\documentclass[11pt,a4paper]{article}
\usepackage{jinstpub}
\usepackage[symbol]{footmisc}        
\usepackage{lineno}
  
\newcommand{\okina}{\textquotesingle}  

\title{Fluorescence time profile measurement of LAB based liquid scintillator in response to medium relativistic ion particles}

\author[a,~b]{Xiaojie Luo}
\author[c]{, Shuya Jin}
\author[a,~*]{, Gaosong Li~\note{Corresponding author: ligs@ihep.ac.cn}}
\author[d]{, Zepeng Li}
\author[e]{, Fenhua Lu}
\author[c]{, Yazhou Sun}
\author[c]{, Shitao Wang}
\author[f]{, Yaoguang Wang}
\author[a]{, Yifang Wang}
\author[g]{, Xiaobao Wei}
\author[a,~\dagger]{, Liangjian Wen~\note{Corresponding author: wenlj@ihep.ac.cn}}

\affiliation[a]{Institute of High Energy Physics, Chinese Academy of Sciences,\\
  19B Yuquan Road, Beijing 100049, China}
\affiliation[b]{School of Physical Sciences, University of Chinese Academy of Sciences,\\
  19A Yuquan Road, Beijing 100049, China}
\affiliation[c]{Institute of Modern Physics, Chinese Academy of Sciences, \\
509 Nanchang Road, Lanzhou 730000, China}
\affiliation[d]{Department of Physics and Astronomy, University of Hawai\okina i at M\={a}noa,\\ Honolulu, HI 96822, USA}
\affiliation[e]{Department of Physics, Southern University of Science and Technology,\\ 1088 Xueyuan Road, Shenzhen 
518055, China}
\affiliation[f]{School of Physics, Shandong University,\\27 Shanda Nanlu, Jinan 250100, China}
\affiliation[g]{School of Physics, Henan Normal University,\\ 46 Jianshe Road East, Xinxiang 453007, China}

\emailAdd{ligs@ihep.ac.cn}
\abstract{Liquid scintillator is widely used in particle physics experiments due to its high light yield, good timing resolution, scalability and low cost. Certain liquid scintillators exhibit pulse shape discrimination capabilities because of difference in fluorescence timing properties induced by different particles. Its fluoresence timing properties have been measured mostly for radioactive decay sources at MeV energies. We present a novel measurement of fluorescence time properties of LAB based liquid scintillator in response to high-energy ions of hydrogen (Z = 1), helium (Z = 2) and Krypton at around 200-300 MeV/u for the first time. We compared the results to those from radioactive sources and observed a distinct $dE/dX$ dependence, regardless of the particle type. These findings are essential for physics searches such as the diffuse supernova neutrino background in large liquid scintillator detectors like JUNO, and are also critical towards understanding the underlying scintillation timing mechanism.}

\begin{document}
\maketitle
\flushbottom

%\linenumbers 

%=================%
\section{Introduction}
\label{sec:introduction}
Organic scintillator technology has been developed since the 1950s and widely used in particle physics experiments. In particular, organic liquid scintillators (LS) have been utilized comprehensively for large detectors in neutrino physics to study neutrino oscillation \cite{kamland2002,dyb2012,dc2013,reno}, solar neutrinos \cite{BX}, neutrinoless double beta decay \cite{kz2022,sno+} among others, given their advantages of low cost, low energy threshold, high purity, homogeneity, and scalability. 

The energy response to different particles has been studied and various models are developed to describe the quenching effect in organic scintillators. Birks first proposed an empirical model to describe the fluorescence light intensity per path length ($dL/dX$) as a function of ionization energy loss per path length ($dE/dX$) in the scintillator \cite{Birks:1951boa}. Chou and Wright expanded it to a more generalized form by taking higher-order dependence on $dE/dX$ into the model \cite{Chou:1952jqv,Wright1953}. Other more complicated models are proposed to model the response at a microscopic level \cite{Voltz1966,Ahlen1977}. Extensive measurements of responses to different ions have been studied throughout the years \cite{BECCHETTI197693, FOX199663,POSCHL2021164865}. 

The timing response is not as thoroughly studied. It was found that organic scintillators exhibit different decay time behaviors for electrons and $\alpha$ particles from radioactive decays \cite{Wright1956}. This is followed by numerous studies and used widely for particle identification applications between neutron/alpha and e/gamma \cite{MOSZYNSKI1979}. More recent research over the past two decades has characterized the decay time constants of different types, or mixtures of LS in response to alpha/beta particles in more detail \cite{MarrodanUndagoitia:2009k,Li2011TimingPA,Lombardi:2013nla}. However, very limited studies are found beyond using particles at the MeV region. 

The upcoming Jiangmen Underground Neutrino Observatory (JUNO) experiment, with 20 kilotons of liquid scintillator as target mass, is going to be the largest scintillator detector \cite{juno_YB,juno_ppnp}. The large target mass significantly expands its physics potential, which will have world-leading sensitivity to diffuse supernova neutrino background (DSNB) \cite{juno_dsnb}. In the DSNB search, the dominant background is atmospheric neutrino neutral-current interactions. Pulse Shape Discrimination (PSD) is critical to differentiate these background events, usually involving neutrons, from inverse beta decay produced by DSNB interactions \cite{dsnb_psd}. This requires a better understanding of the scintillation time response for high-energy neutrons and protons at the level of tens of MeV. The energy and time response are well motivated to be driven by the same physics; a combined understanding of scintillation energy and time response could also help improve the understanding of the energy resolution \cite{juno_resolution}, which is the key for the neutrino mass ordering (NMO) measurement with reactor neutrinos \cite{juno_reactor_nmo}. 

In this work, we use an ion beam with different types of ions to measure the liquid scintillation time response to high-energy particles and fragments for the first time. This represents a first attempt towards a unified time model.

The rest of the paper is organized as follows: In Sec.~\ref{sec:exp}, we will introduce the overall experimental design; the data and analysis details  are described in Sec.~\ref{sec:ana}; the results of the timing responses for different particles, systematical uncertainty evaluations, and discussions are discussed in Sec.~\ref{sec:res}; finally we conclude in Sec.~\ref{sec:conclusion}.

%=========================%
\section{Experimental methods}
\label{sec:exp}

Typical fluorescence time measurements in LS employ electron/$\gamma$ sources and n/$\alpha$ sources at energies of several MeV, corresponding to ionization densities $dE/dX$ from $\sim$0.2~MeV/mm for (e, $\gamma$) and $\sim$130~MeV/mm for $\alpha$. The various fluorescence time constants of $\gamma$/$\alpha$/neutron are generally believed  to be due to their different ionization densities
in the scintillator. It would be important to directly measure and compare the time spectrum with different particles of similar $dE/dX$. Furthermore, to fully characterize the timing response, it is essential to obtain data spanning the intermediate $dE/dX$ region. Either lower energy electron sources or higher energy heavier particles can provide $dE/dX$ of interests. In this work, we designed a detector setup to measure the timing response for high-energy ion beam particles through a thin slice of LS. The details about the experimental methods are described in the following.

\subsection{Beam data}
\label{sec:Beam}
\begin{figure}
\centering
\includegraphics[width=1\linewidth]{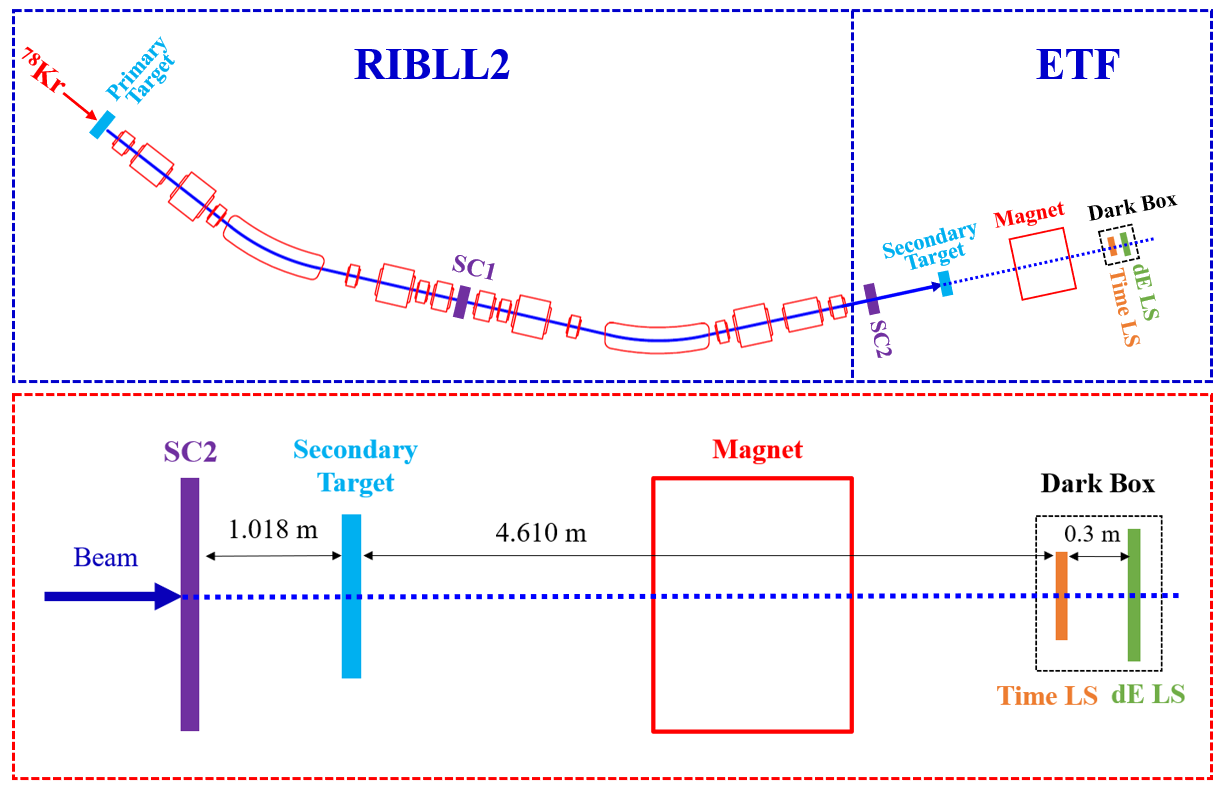}
\caption{\label{fig:Experimental Layout}  Schematic layout of the experiment. The upper panels show the full view of the RIBLL2 beamline and the External Target Facility (ETF), showing the plastic scintillators~(SC1 and SC2) and the optional targets~(see Table~\ref{tab:beam}). The bottom panel presents a zoom of ETF region, highlighting the elements used in this study. The downstream plastic scintillator~(SC2) and dE liquid scintillator~(LS) provide the start and the end signals for time of flight in this research. The LS samples to be measured are located at the dark box region whose details are shown in Figure~\ref{fig:Experimental Setup in Dark Box}.} 
\end{figure}

The ion beams are supplied by the Radioactive Ion Beam Line in Lanzhou-II (RIBLL2) \citep{SUN201878, Tang2012} at the Cooler Storage Ring of the Heavy Ion Research Facility in Lanzhou (HIRFL-CSR) \citep{XIA200211}. Our detector is located at the External Target Facility (ETF) \citep{SUN201872, SUN2019390} area. The schematic experimental layout is shown in Figure~\ref{fig:Experimental Layout}. The primary or heavy secondary beams from RIBLL2 first traversed a \textbf{plastic scintillator (SC2)}, which provides the start signal for \textbf{time-of-flight (TOF)} measurements. The beams then impinged on a secondary target, where light fragments were produced. A dipole magnet downstream of the target can redirect the beam to a separate experimental area. In this work, our LS detector is positioned along the primary beam axis.

For this study, data were collected under two beam configurations:

1) \textbf{$^{78}$Kr beam}: No primary or secondary targets were placed in the beam line, and the magnetic field is set to zero. Therefore, the 300 MeV/u $^{78}$Kr beams pass directly to the LS targets. 

2) \textbf{Light Fragments}: A 350 MeV/u $^{78}$Kr beam struck  a 5-mm beryllium primary target, followed by a carbon (C) target. The magnetic field is set to 0.9 Tesla. So the heavy particles are bent away. Residual off-axis fragments (mostly  Z=1 and Z=2, where Z denotes atomic number) are left to timing spectrum measurement. The used beam configurations are summarized in Table~\ref{tab:beam}.

\begin{table}[htbp]
\centering
\begin{tabular}{l|llll}
\hline
Dataset & Primary Target & Secondary Target & Magnet field & Particles through LS \\
\hline
Kr Beam& no & no & off & $^{78}$Kr \\
Light Fragments &  Be & C & on & light particles\\
\hline
\end{tabular}
\caption{Beam data configurations.}
\label{tab:beam}
\end{table}

\subsection{Detector design}
\label{sec:det}
\begin{figure}
\centering
\includegraphics[width=1\linewidth]{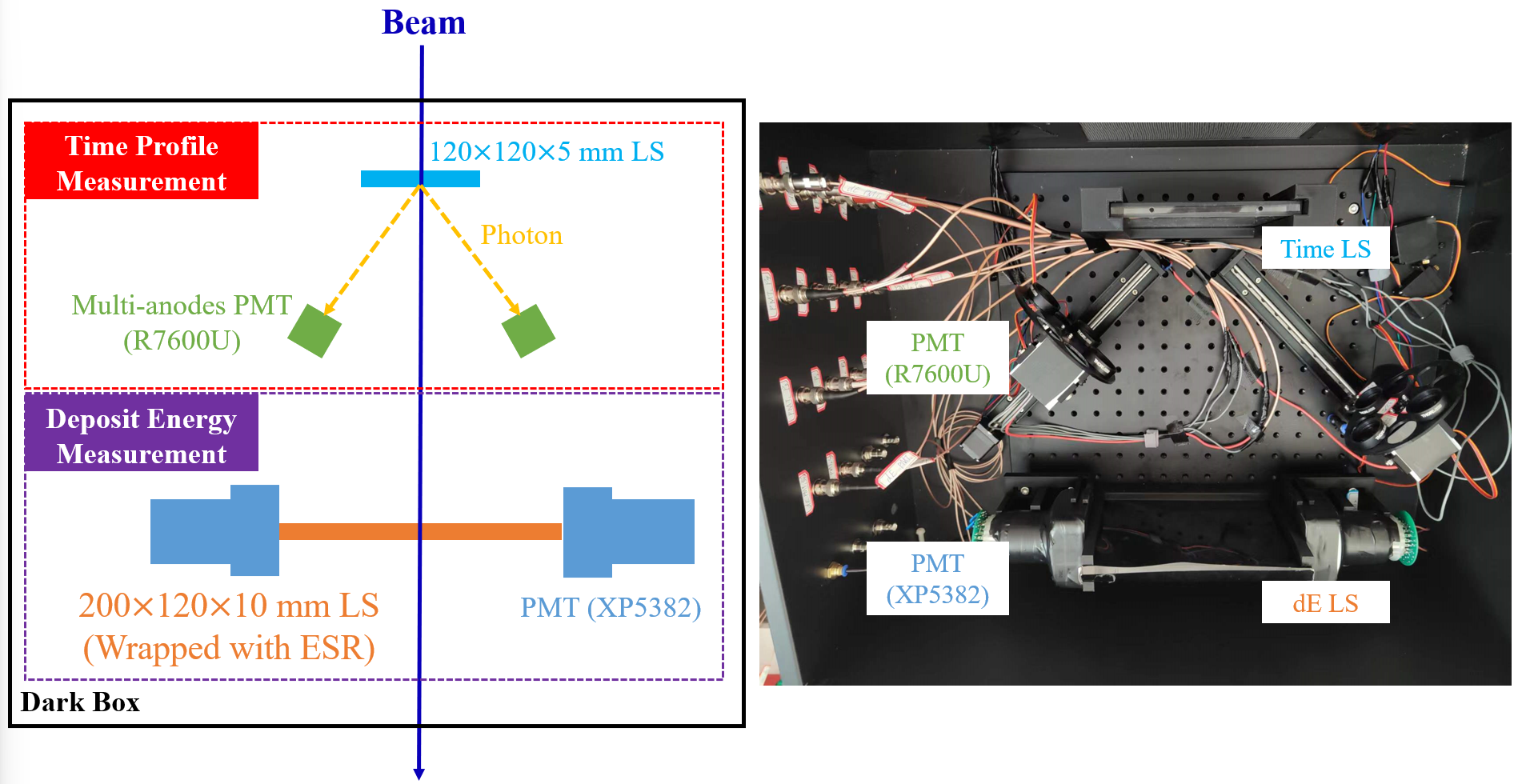}
\caption{ The left panel shows the schematic view of the LS detector in the dark box, and the right panel is the photo of the setup. The detector contains two parts, including time profile and deposit energy measurement. }
\label{fig:Experimental Setup in Dark Box}
\end{figure}

The LS samples are housed in a dark box with dimensions of 60~cm~$\times$~38~cm~$\times$~23~cm. The experimental configuration inside the dark box is illustrated in Figure~\ref{fig:Experimental Setup in Dark Box}. Its inner surface is coated black to minimize light reflection, and a 200~mm $\times$ 150~mm aperture covered with 0.2~mm light-blocking fabric tape was integrated to mitigate beam energy deposition before LS targets while excluding ambient light. A 5~mm-thick slice of the LS sample is contained in an acrylic cuvette (12~cm~$\times$~12~cm internal dimensions, 1~mm wall thickness) with a black backing to suppress the internal optical reflections during the time profile measurement. The LS is prepared using a mixture of 2.5 g/L PPO and 3 mg/L Bis-MSB dissolved in Linear Alkyl Benzene (LAB). Two fast multi-anode PMTs (R7600U, HAMAMATSU) are placed after the sample, looking at an angle around 45$^\circ$ relative to the beam direction, with the distance to the sample adjustable. Each PMT has four independent channels arranged in a 2$\times$2 array. In total, eight channels are used to measure the timing spectrum to improve statistics. The timing spectrum is measured in single-photoelectron mode with the delayed coincidence method \cite{Bollinger1961}. The distance between these two PMTs and the sample is adjustable using step motors, allowing the online adjustments so that the PMTs are operated at the expected occupancy. Additional absorptive neutral-density filters (Thorlabs NE20A, NE13A, and NE05A) can be controlled by servomotors\footnote{A servomotor is a compact electromechanical actuator that interprets pulse-width-modulated inputs to hold or move to a specified angular position.} to provide further optical attenuation to extend the dynamic range. An additional aperture in different sizes in front of individual detection channels provides further handle on the occupancy adjustment to accommodate the particles with significantly differed $dE/dX$. The setup is designed to be capable of measuring beams with incoming particles $dE/dX$ differing by 1-2 orders of magnitude.

A larger LS sample of the same formulation (20~cm~$\times$~12~cm~$\times$~10~mm) was placed downstream of the timing sample as a transmission detector for deposit-energy measurement. This sample is fully wrapped by an Enhanced Specular Reflector (ESR) film to increase light-collection efficiency. Dual-ended readout is performed by two $2^{\prime\prime}$ dynode PMTs (XP-5382), referred as dE-PMTs, with 6~cm~$\times$~1.2~cm apertures. The time difference between the plastic scintillator and these dE-PMTs gives the time of flight for particles. Combined with the deposited energy in this sample, particle identification can be performed. Additionally, this LS sample enables the accurate determination of the starting time for the fluorescence time measurement.

The controls of motors, PMT high voltages, and data taking, are operated remotely during beam time. Measurements are performed at room temperature ($16.7 \pm 0.5^\circ\mathrm{C}$). To mitigate the oxygen quenching effect \cite{PhysRev.101.998,XiaoHua-Lin_2010,Li2011TimingPA}, the LS samples are purged with nitrogen and sealed with a fluorine rubber plug before measurement. Additionally, an oxygen content meter is installed inside the dark box to monitor the oxygen concentration. The dark box is purged with nitrogen at a flow rate of 2.0 L/min read out by a flowmeter to maintain the oxygen concentration inside at approximately 0.01\% during the measurements.

\subsection{Electronics and Data acquisition}
A CAEN 751 digitizer family (1 GS/s) is employed for waveform readout in this experiment, utilizing a 4 µs window length for each event. To eliminate the jitter caused by the digitizers (the 125 MHz clock cycle of the 751 series induces an 8 ns jitter), the waveform of the trigger source is also recorded to ensure precise time measurement.

In the trigger perspective, two beams were adapted to different trigger sources. The primary beam utilizes the plastic scintillator signal ($T_\mathrm{stop}$) as the trigger source, whereas the fragments employ the coincidence of two PMTs in the dE measurement part. This modification is necessary because most of the data triggered by $T_\mathrm{stop}$ are empty, while only approximately 1\% of beam fragments reach the LS detector.

%===============================%
\section{Measurement and Data Analysis}
\label{sec:ana}

\subsection{Z=1, Z=2 beam data}\label{subsec:beam_data}
A beam of $^{78}$Kr ions (350 MeV/u) collides with a Be target, resulting in beams consisting of mostly nuclei with A/Z$\sim$2. The beam travels around 20 meters before colliding with a secondary target. A magnetic field ($\sim$0.9 tesla)  bends the fragmented nuclei away from our experimental zone. The majority of forward-traveling nuclei are deflected by the magnetic field away from the LS target, while nuclei with large scattering angles from the collision with the secondary target can be bent back by the field and registered on the LS sample. These particles are dominantly low-mass and have Z=1 or 2.

In this beam configuration, the trigger is issued by the coincidence of the two dE-PMTs. The $T_\mathrm{stop}$ signal provided by the plastic scintillator is also recorded so that the TOF from the plastic scintillator to the dE-PMTs can be calculated, providing the velocity of the nuclei (or kinetic energy per nucleon, MeV/u). The total energy ($\Delta E$) deposited in the LS sample is measured by the two dE-PMTs. The $\Delta E$-TOF method is used to identify the fragment particles \cite{FOOT2019,POSCHL2021164865}. According to Bethe-Block's theorem \cite{Bethe:1930ku}, $\Delta E$ of charged particles is proportional to the square of the atomic number (Z$^2$). Combining particle velocities derived from the measured TOF,  one can achieve identification of particles with different atomic numbers $Z$.

\begin{figure}
\centering
\includegraphics[width=1\linewidth]{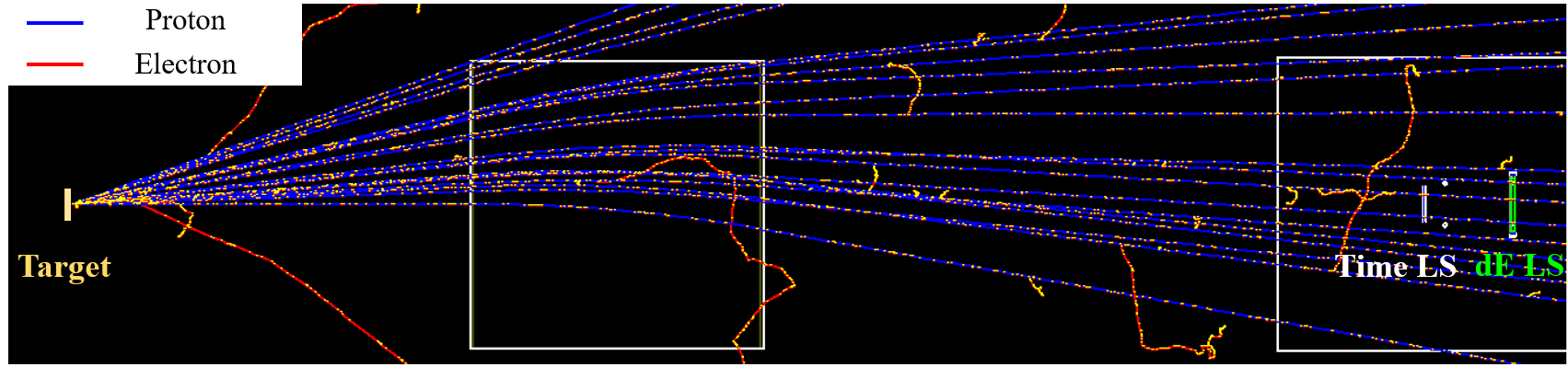}
\caption{\label{fig:magnetic_sim}  Visualization of the Geant4 simulation of the secondary particles that could pass through the dE LS sample. Protons with varying energies were ejected from different angles ($0^\mathrm{o}-21^\mathrm{o}$) and the magnetic field was set as 0.9~Tesla in accordance with the actual scenario. The blue lines represent the simulated tracks of protons. Particles within a certain range of scattering angles can be directed to the LS sample. }
\end{figure}

A GEANT4 simulation  is built to compare with the data. Nuclei with Z=1,2,3 emitted after the secondary target are simulated. Particles are generated with a uniform energy distribution from 0 to 1000 MeV/u from the target. The magnetic field is configured to be uniform with a square-shaped region of 0.9 Tesla strength in the simulation. The geometry is shown in Figure~\ref{fig:magnetic_sim}. A wavelength-dependent reflectivity for the Enhanced Specular Reflector (ESR) is used, with the average value of 98\% from 380 to 750 nm \cite{An_2016_ESR}. No dependence of reflectivity on the angle of incidence and polarization is implemented. The quantum efficiency of the dE-PMTs is taken from the vendor's manual\footnote{https://hallcweb.jlab.org/DocDB/0008/000809/001/PhotonisCatalog.pdf}. The optical process considers the photon absorption and re-emission \cite{lin_tao_simulation_2023,yang_research_2024}. The simulation is benchmarked with a measurement of the 59.5 keV gammas from an $^{241}$Am source. 

\begin{figure}
\centering
\includegraphics[width=0.6\linewidth]{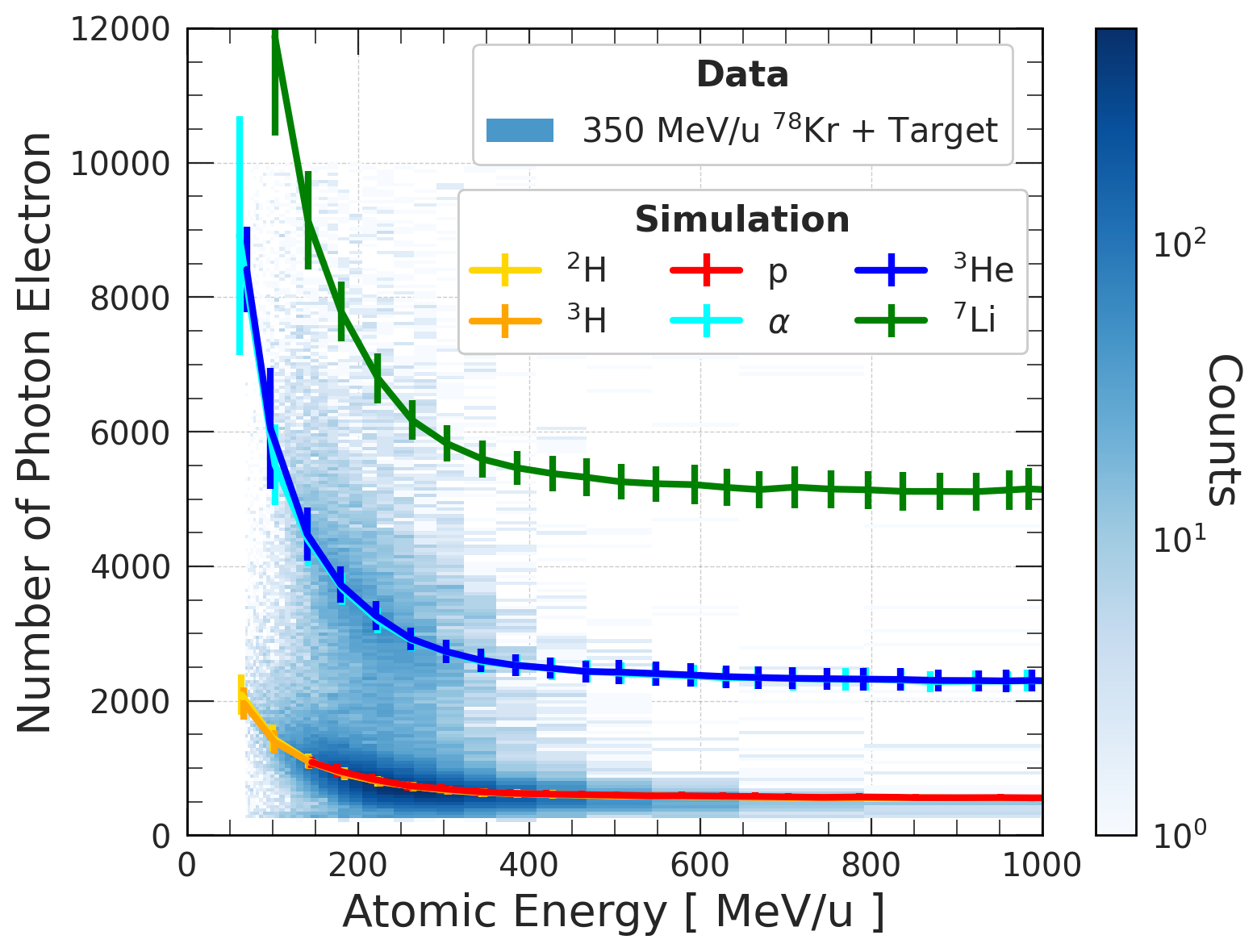}
\caption{\label{fig:dEvsE_dataMC} $\Delta E$ vs. E distribution. The 2D histogram represents the data, while the lines are medians from MC predictions for Z=1,2,3 particles. The two prominent clusters in the data align with the median of Z=1 and Z=2 particles in MC. }
\end{figure}

The energy deposition in the LS sample vs atomic kinetic energy is shown in Figure~\ref{fig:dEvsE_dataMC}. The gain of the two dE-PMTs was calibrated by low-intensity LED data. Two clusters of population can be clearly observed.  The MC predicts nuclei with the same atomic number Z follow a cluster. The distribution is in good agreement with data. A clear separation between nuclei with different Z is observed. A cut of $\pm$50\% within the MC median is used to select samples of Z=1 and Z=2 particles. 

\subsubsection{Event selection}
Multiple particles from the beam may enter a single readout window, potentially biasing the time distribution. We require that only one signal be detected in the $T_\mathrm{stop}$ signal. This requirement excludes 13\% of the total events.

\subsubsection{Hit selection}
\label{sec:hit_sel}
\begin{figure}
\centering
\includegraphics[width=0.6\linewidth]{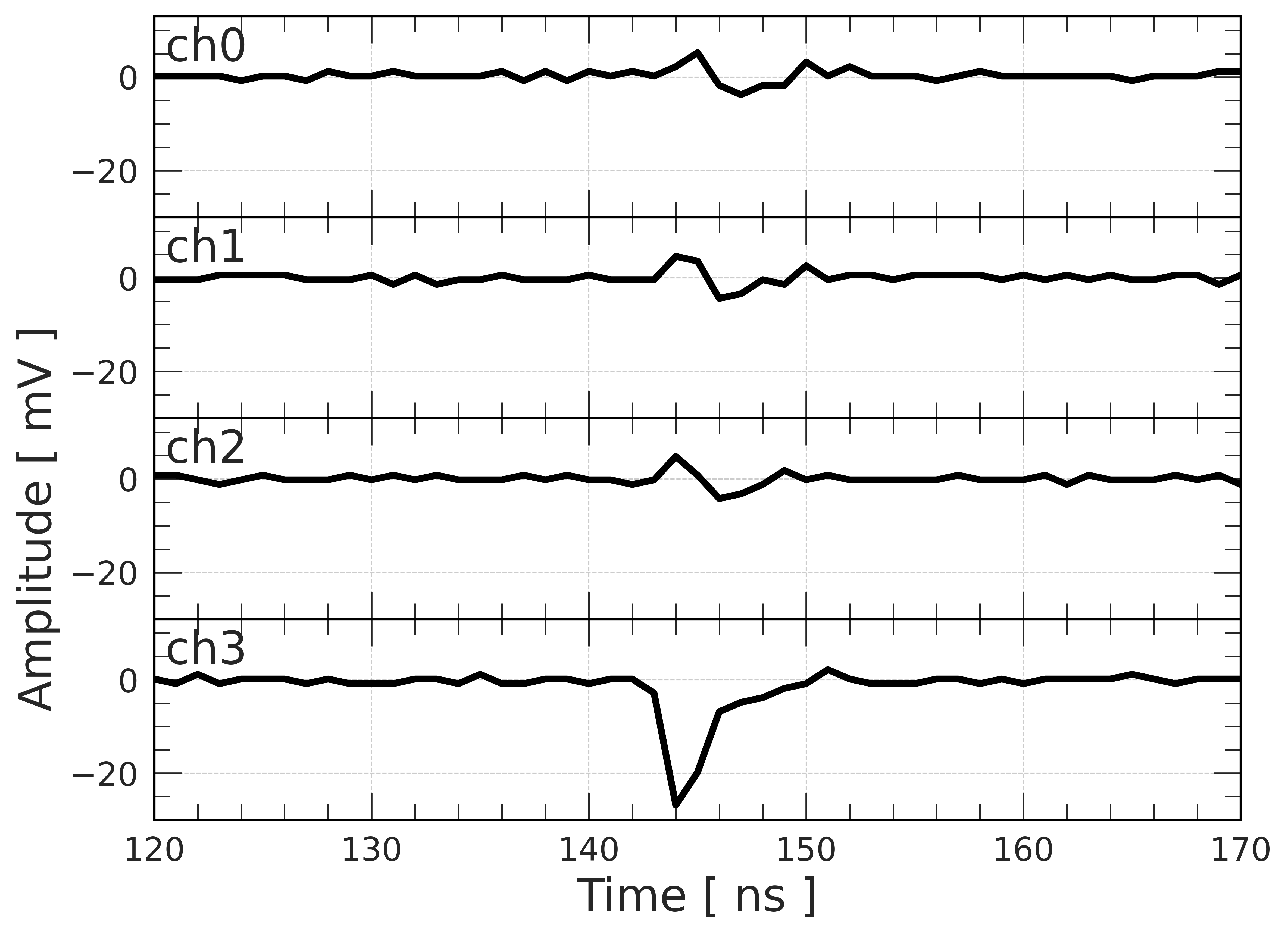}
\caption{\label{fig:cross_talk} Single-event waveform from all channels of an R7600U PMT. The bottom channel (ch3) serves as the signal channel, while the other channels exhibit crosstalk noise at roughly 20\% of the signal amplitude and occur simultaneously with the signal. }
\end{figure}

\textbf{Cross talk cut} The timing PMT R7600U features a multi-anode design. According to the vendor's manual, a $\sim$1\% cross talk probability is observed for adjacent channels and the probability is smaller for the diagonal channel. This is expected to have a negligible impact on the measured timing distribution. Another type of electrical cross talk is observed in the data. The electrical cross talk waveform shows a bipolar behavior, very distinctive from a true photon signal as shown in Figure~\ref{fig:cross_talk}. The amplitude of the cross talk is around 10-20\% of the primary signal. Such electrical cross talk signals are studied by illuminating one channel with the other three masked. The cross talks can be identified with high efficiency (>99.9\%) while retaining 92.3\% efficiency for physical signals. The signal-to-background ratio can be improved by a factor of $10^3$. On certain occasions, two channels are hit simultaneously; the overlap of the true photon signal and the cross talk signal from the other channel could lead to a bias or inefficiency of the signal. The impact is studied and found to be marginal, which will be discussed in Sec.~\ref{sec:syst}.

\noindent\textbf{Single hit cut} A requirement for a single hit on each timing channel is implemented. This ensures the measured time spectrum remains unbiased. Hits that are close (within the time resolution) together are challenging to separate, which results in the suppression of the fast component fraction within several tens of nanoseconds. The occurrence of afterpulses following a real single-hit event introduces a slight signal inefficiency, leading to a $\sim$5\% signal loss by this requirement.

\noindent\textbf{Single pe cut} With the single-hit requirement in place, hits that are close together ($\Delta T< 10~ns$) cannot be distinguished. They are reconstructed as one hit, leading to an underestimation of the fast component. A single photoelectron (pe) cut is further applied to remove multiple-pe events. If a pulse has charge greater than 1.5 pe, the event is excluded. This leads to a signal efficiency of 95\% while rejecting more than half of the two-pe events. The impact on efficiency and purity is studied using the realistic single photoelectron (SPE) response of the R7600U PMT. The SPE response is derived from a low-intensity LED run. The potential impact is detailed in Sec.~\ref{sec:syst}.

\subsection{Krypton beam data}
The main Krypton beam, at 300 MeV/u, is directed straight into the experimental area without any Be or C targets in its path. No magnetic field was applied. The trigger signal is produced by the $T_\mathrm{stop}$ as registered by Kr in the plastic scintillator. The same event and hit selection criteria as above were applied.

\subsection{Radioactive sources}
We also measured the time distributions using radioactive sources: 5.5 MeV alpha particles from $^{241}$Am and electrons from $^{207}$Bi. The setup described in Sec.~\ref{sec:det} was utilized with minor adjustments. To facilitate interactions of those low-energy electrons and alpha particles with LS while suppressing the impact from the radioactivity background, an open-topped and smaller (2~cm~$\times$~2~cm, 5~mm thickness) container, similar in design to the first LS sample, was adapted. The radioactive sources were positioned on top of the acrylic container, while an additional 2$''$ PMT (R13435, HAMAMATSU) with fast time response was placed on the side of the sample to provide both the trigger signal and the start time for the fluorescence time measurement. The same hit selection criteria as outlined in Sec.~\ref{sec:hit_sel} were applied.

\subsection{Instrumental time response}
\begin{figure}
\centering
\includegraphics[width=0.6\linewidth]{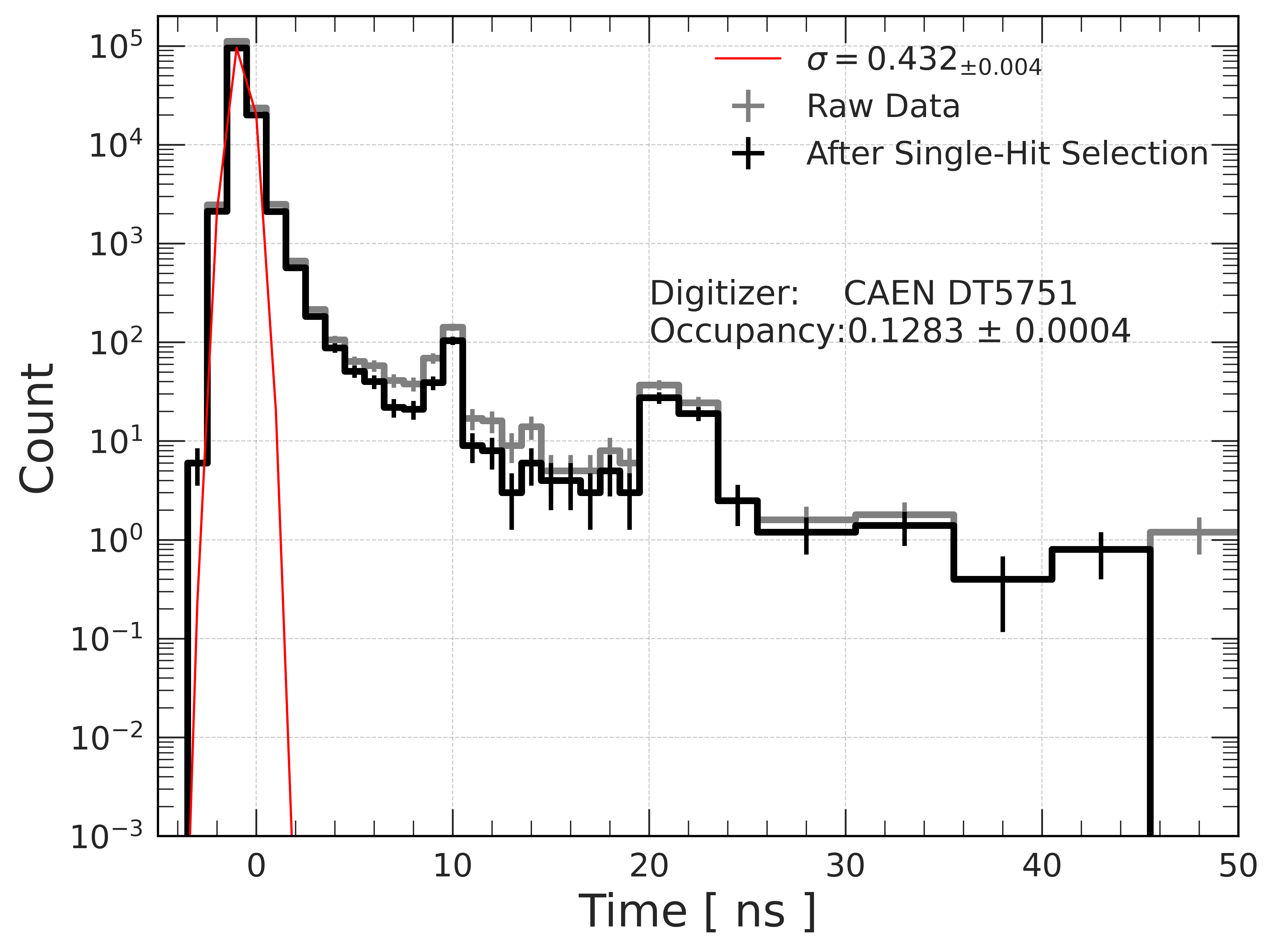}
\caption{\label{fig:time_response}  The detector’s temporal response was measured using a picosecond laser. The peak exhibits a timing resolution of approximately 0.4 ns. The gray histogram corresponds to the raw dataset, where all recorded hits are included, while the black histogram reflects events filtered to retain only a single hit per trigger. }
\end{figure}

The time response of the detector system is studied using a PICO P-C-405 laser with a 20 ps full width at half maximum (FWHM) at 405 nm. Data readout is initiated by an external trigger that also drives the laser. The laser operates at a low intensity, allowing the R7600U PMTs to detect single photoelectrons most of the time. The occupancy is approximately 0.2, which closely matches the conditions of the timing measurement. The same hit selection procedure in Sec.~\ref{sec:hit_sel} was also applied.The measured response is shown in Figure~\ref{fig:time_response}. The main Gaussian peak gives a timing resolution of 0.41 ns. A minor non-gaussian tail is observed. The shape of the instrumental response is fully taken into account to extract timing constants in Sec.~\ref{sec:res}. The impact is negligible though due to the small contribution of the tail.

%================================%
\section{Time Spectrum Results}
\label{sec:res}
As described in the previous sections, we have obtained time spectra for five different particles, with the key information summarized in Table~\ref{tab:info}. The particles vary in type and energy. Due to the constraints on beam time, the total statistics for Z=1 and Z=2 particles are lower by an order of magnitude compared to the others, resulting in larger statistical variations in the measured spectra.

\begin{figure}
\centering
\includegraphics[width=1.0\linewidth]{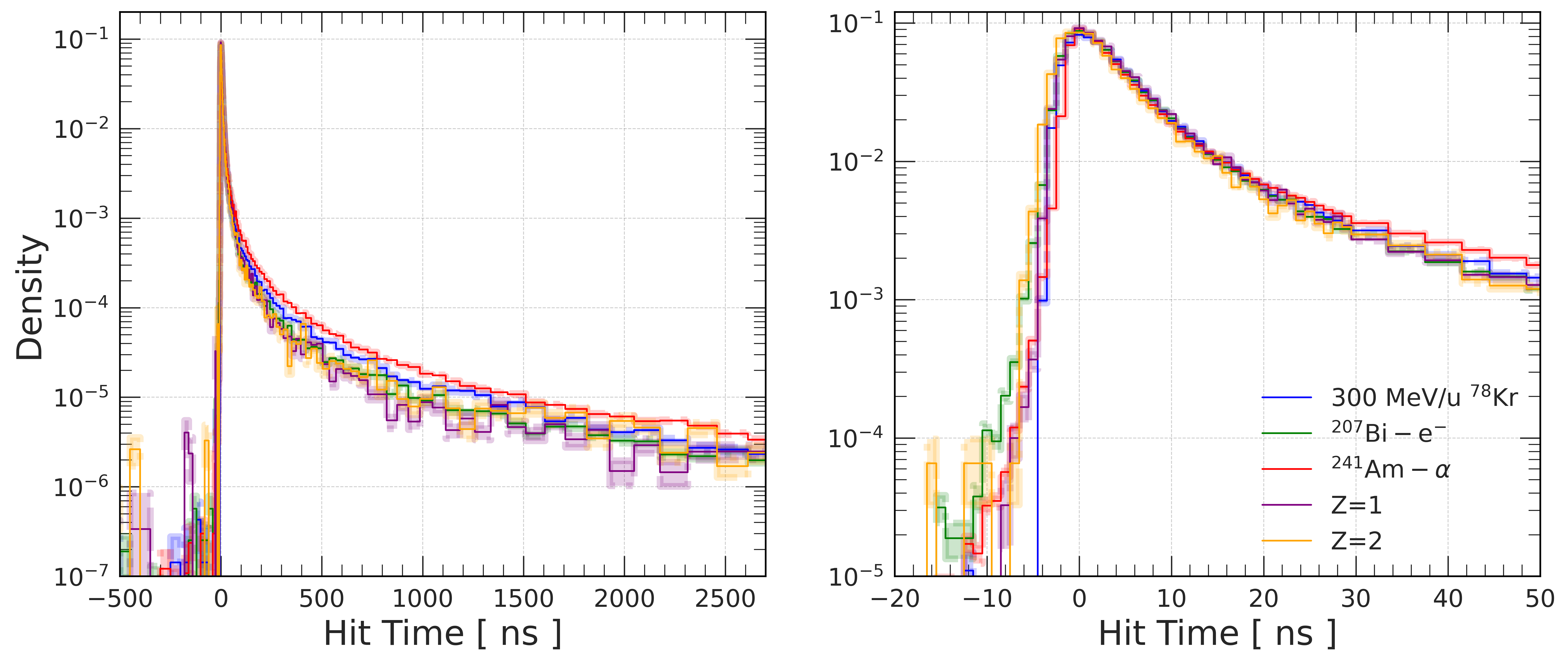}
\caption{\label{fig:time_spectra} The measured scintillation time spectra (left) and a zoom-in to the peak region (right). The shaded bands show the statistical uncertainty. The dark noise contribution has been subtracted. }
\end{figure}

The final time spectra are plotted in Figure~\ref{fig:time_spectra}. A clear difference in the fraction of the delayed time components between alpha and electrons from radioactive sources is observed, in agreement with measurements by \cite{Beretta:2025rwj}. This allows the identification of alpha/beta particles in the MeV region. The curves for Z=1 and Z=2 particles are very close to that of electrons, despite slightly insufficient statistics at longer time. The $^{78}$Kr data lies between the electron and alpha. Discussions on this behavior are detailed in Sec.~\ref{sec:discussion}.

\begin{table}[htbp]
\centering
\begin{tabular}{llll}
\hline
Particle Type & $dE/dX$ [MeV/mm] & Kinetic Energy & Statistics \\
\hline
$^{78}$Kr & 357$\pm$2 & 300 MeV/u & 175,000 \\
Z=1 & 0.30$\pm$0.06 & 209$\pm$73 MeV/u & 30,000 \\
Z=2 & 1.30$\pm$0.29 & 188$\pm$60 MeV/u & 15,000 \\
$^{207}\text{Bi}: \text{e}^-$ &  0.19$\pm$0.03  & $\sim$1 MeV &  158,000\\
$^{241}\text{Am}: \alpha$ & 127$\pm$2 & 5.5 MeV &  786,000\\
\hline
\end{tabular}
\caption{Summary of the different sources. $dE$/$dX$ value is based on simulation, accounting for the energy deposition from the main track.}
\label{tab:info}
\end{table}

\subsection{Time constants}
To extract the decay time constants, the time profiles are fit with a four-component exponential function. To consider the detector response, the timing response of R7600U PMT ($\mathrm{R}_\mathrm{PMT}(t)$) shown in Figure \ref{fig:time_response} is convoluted with the intrinsic scintillation function. In addition,  an extra Gaussian smearing ($\mathrm{G}(t)$) is integrated. The Dark Noise (DN) of PMT is described with a fixed constant determined by getting the mean value of the $\Delta T<0$ region. The time profile is parameterized as below:

\begin{equation}
    f(t) = \mathrm{C}_\mathrm{DN} + \mathrm{G}(t) \otimes \mathrm{R}_{\mathrm{PMT}}(t) \otimes \sum_{i=1,2,3,4} A_i \left( e^{-\frac{t-{t_0}}{\tau_i}} - e^{-\frac{t - t_0}{\tau_r}} \right)
    \label{eq:fitting_model}
\end{equation}
\[
\begin{aligned}
    &\tau_i  && \text{Decay time constants} \\
    &A_i  && \text{Weights of decay components} \\
    &\tau_r  && \text{Time constant for rising edge} \\
    &\mathrm{C}_\mathrm{DN}  && \text{Constant for PMT dark noise} \\
    &\mathrm{G}(t) && \text{Gaussian distribution for additional detector resolution} \\
    &\mathrm{R}_{\mathrm{PMT}}(t)  && \text{PMT response function (from laser experiment)} \\
    &t_0  && \text{Time offset of peak}
\end{aligned}
\]

\begin{figure}
\centering
\includegraphics[width=1\linewidth]{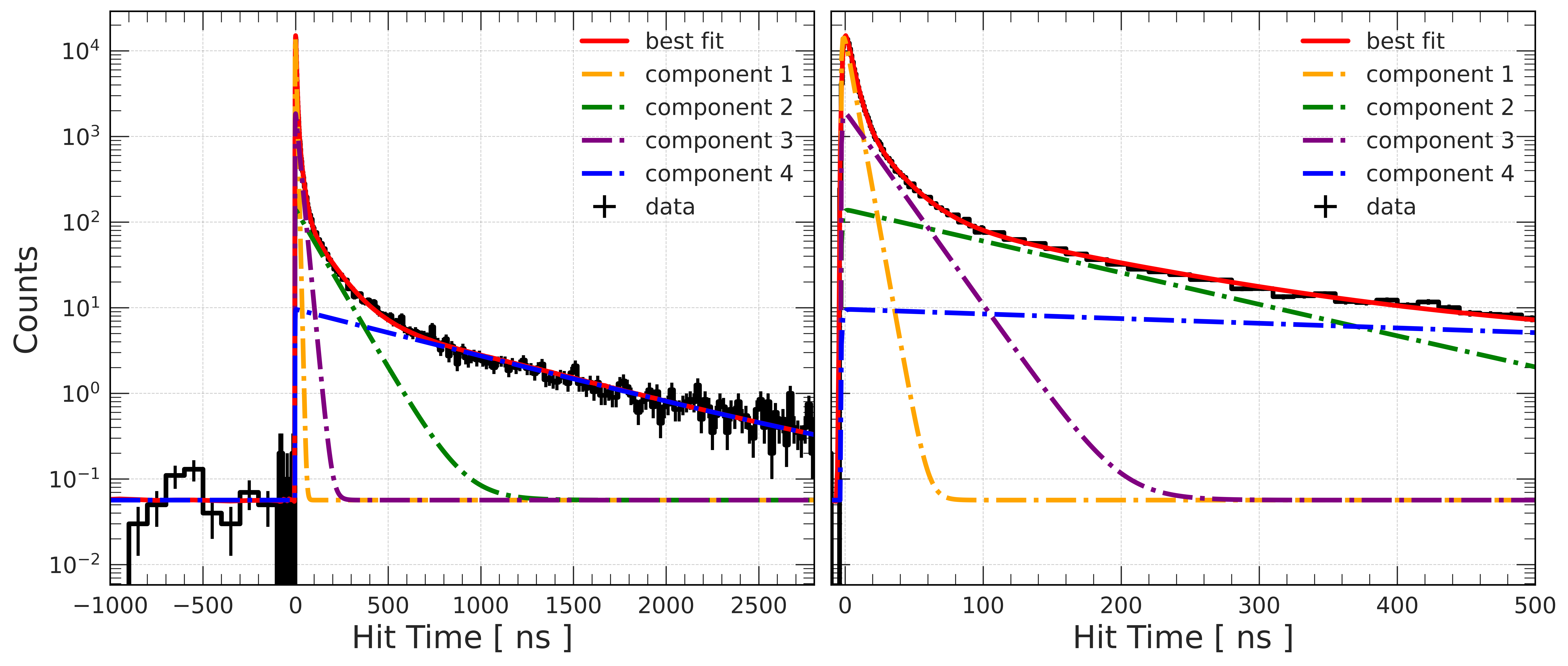}
\caption{\label{fig:best_fit}  An example best fit to the time spectrum of 300~MeV/u $^{78}$Kr (left) and the zoom-in to peak region (right). }
\end{figure}

\begin{figure}
\centering
\includegraphics[width=0.6\linewidth]{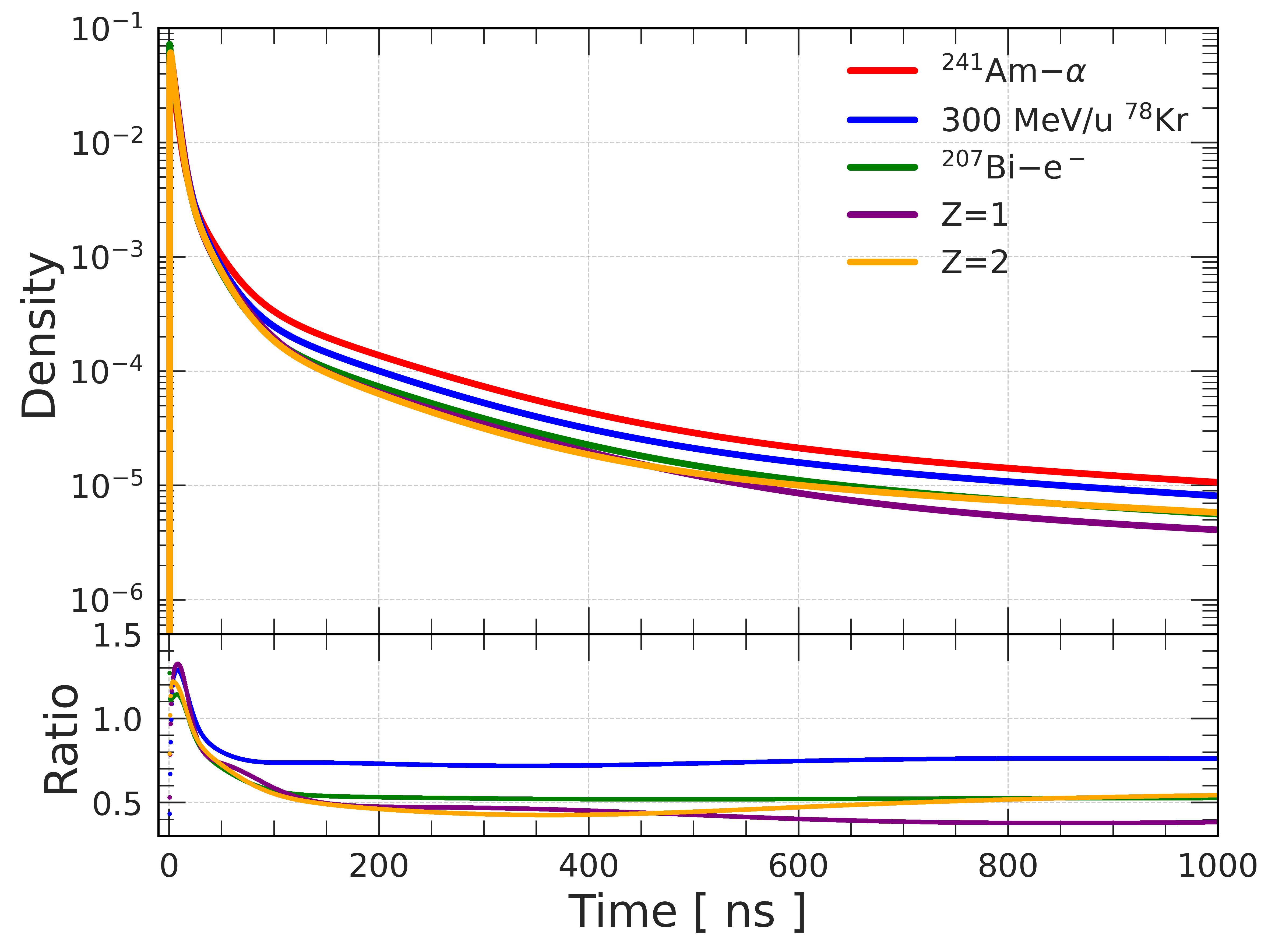}
\caption{\label{fig:deconv_profiles} Comparison of the scintillation time profile using the best-fit time constants and the ratio to $^{241}$Am-$\alpha$. }
\end{figure}

The binned data spectrum is fitted to $f(t)$ with the extended binned maximum-likelihood fit method. 
The time offset of peak $t_0$ is manually scanned during the fit to avoid failed gradient due to rapid change of likelihood. The minimization is done with the Python package \texttt{iminuit}. A best fit to the spectrum of $^{78}$Kr is shown in Figure \ref{fig:best_fit}. The best-fit time constants are summarized in Table~\ref{tab:time_constants}. The best-fit intrinsic time profiles are plotted in Figure \ref{fig:deconv_profiles}. The experiment in this work opened a relatively long readout window (4~$\mu$s) compared to past studies, such that it is sensitive to longer time components. 

\begin{table}[h]
\centering
\begin{tabular}{c|ccccc}
\hline
 & $^{241}$Am-$\alpha$ & 300~MeV/u $^{78}$Kr & $^{207}$Bi-e$^-$ & Z=1 & Z=2 \\
\hline
$\tau_1$ & $4.51 \pm 0.02$ & $4.87 \pm 0.06$ & $4.88 \pm 0.03$ & $5.3 \pm 0.1$ & $4.7 \pm 0.1$ \\
$A_1$ & $0.544 \pm 0.002$ & $0.616 \pm 0.006$ & $0.67 \pm 0.01$ & $0.71 \pm 0.01$ & $0.66 \pm 0.02$ \\
$\tau_2$ & $21.2 \pm 0.2$ & $19.2 \pm 0.5$ & $20.0 \pm 0.4$ & $24 \pm 1$ & $20 \pm 1$ \\
$A_2$ & $0.269 \pm 0.002$ & $0.243 \pm 0.005$ & $0.232 \pm 0.008$ & $0.206 \pm 0.008$ & $0.24 \pm 0.01$ \\
$\tau_3$ & $125 \pm 1$ & $117 \pm 3$ & $122 \pm 4$ & $137 \pm 11$ & $106 \pm 12$ \\
$A_3$ & $0.131 \pm 0.001$ & $0.097 \pm 0.002$ & $0.070 \pm 0.004$ & $0.064 \pm 0.004$ & $0.067 \pm 0.006$ \\
$\tau_4$ & $845 \pm 9$ & $788 \pm 18$ & $845 \pm 26$ & $1150 \pm 147$ & $918 \pm 88$ \\
$A_4$ & $0.056 \pm 0.001$ & $0.044 \pm 0.001$ & $0.029 \pm 0.002$ & $0.021 \pm 0.001$ & $0.031 \pm 0.002$ \\
\hline
\end{tabular}
\caption{Best-fit time constants. The quoted errors are statistical only.}
\label{tab:time_constants}
\end{table}

\subsection{Systematic uncertainties}
\label{sec:syst}
The distribution of scintillation times can be influenced by various factors. The major systematic uncertainty in the extracted scintillation time is caused by the hit selection described in Sec.~\ref{sec:hit_sel}. In this section, we discuss these factors and their impact.

In this study, the occupancy for the beam data ranges from 0.1 to 0.4. For the dataset of $^{78}$Kr and Z=1, the average number of hits $\mu$ is around 0.2. For Z=2, $\mu$ increases to approximately 0.4 due to differences in $dE/dX$ compared to Z=1. Events with only a single hit in the readout window provide an unbiased time spectrum sample. However, the finite time resolution makes closely occurred hits unresolved. This issue is more likely to occur at earlier times, because the scintillation time distribution is peaking at short time. As a result, a suppression of the fast component is anticipated. The single pe cut was used to reduce the bias. The distortion on the time spectrum is assessed using a toy Monte Carlo simulation, which takes properly into account the realistic effect including scintillation time distributions, SPE response, cross talk and waveform reconstruction.

\begin{figure}[htbp]
\centering
\includegraphics[width=.45\textwidth]{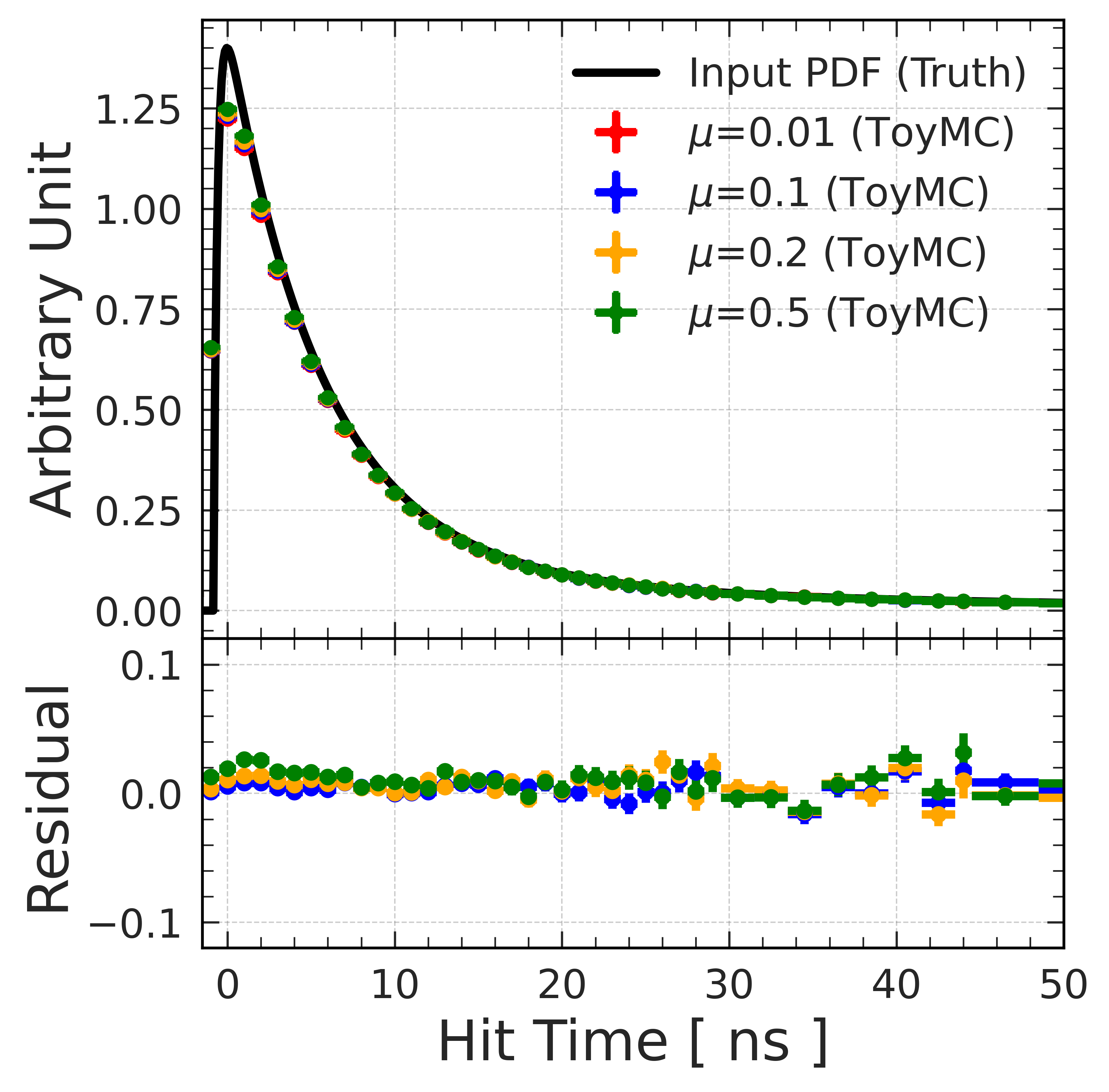}
\qquad
\includegraphics[width=.45\textwidth]{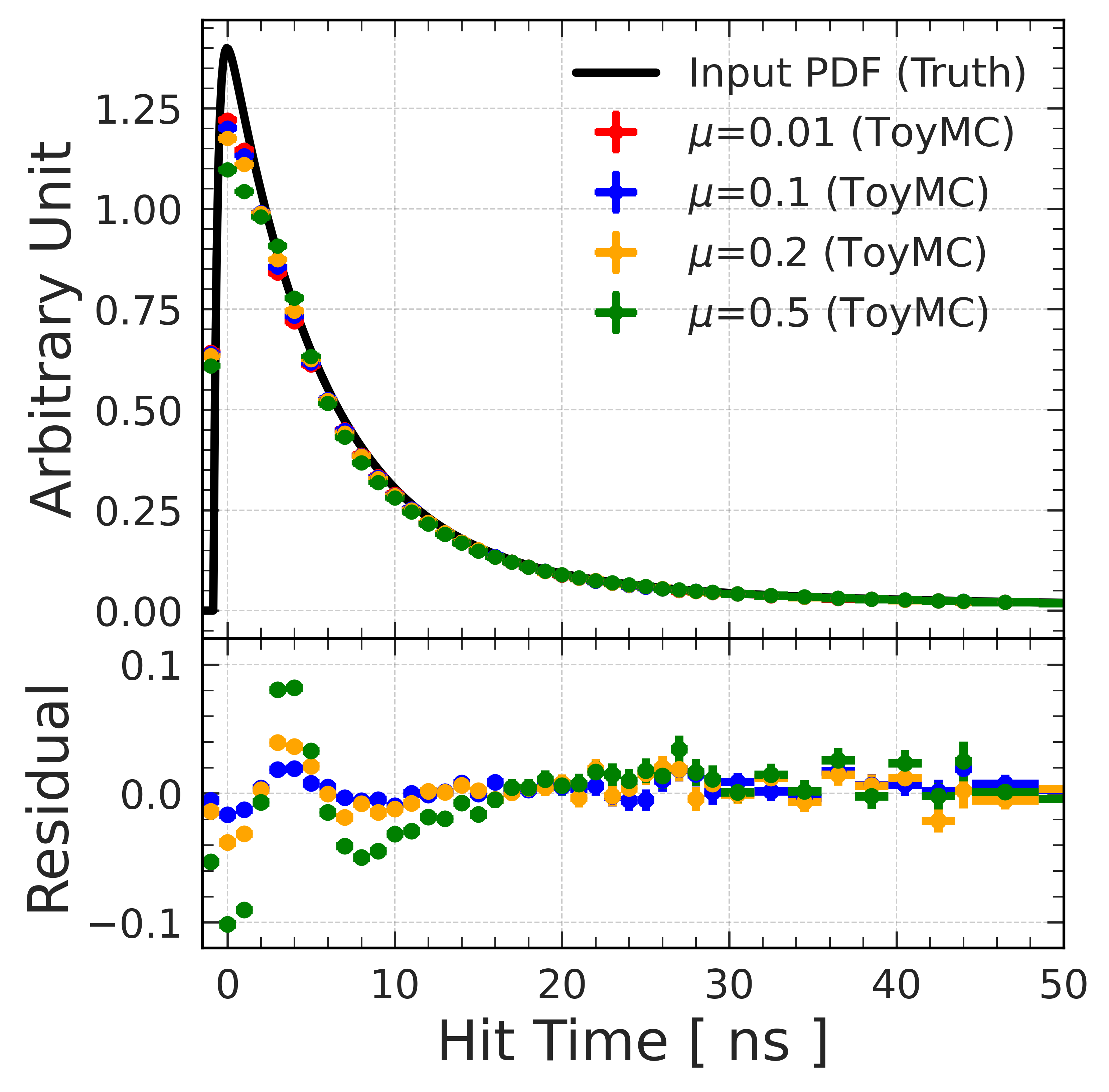}
\caption{Dependence of shape distortion on occupancy. The left panel displays results excluding the cross-talk effect, whereas the right panel includes this effect. Time profiles are normalized by the tail area after 100~ns. The residual represents the relative deviation from the reference time profile at an occupancy of $\mu$=0.01.
\label{fig:occ_dependence}}
\end{figure}

\begin{figure}[htbp]
\centering
\includegraphics[width=.45\textwidth]{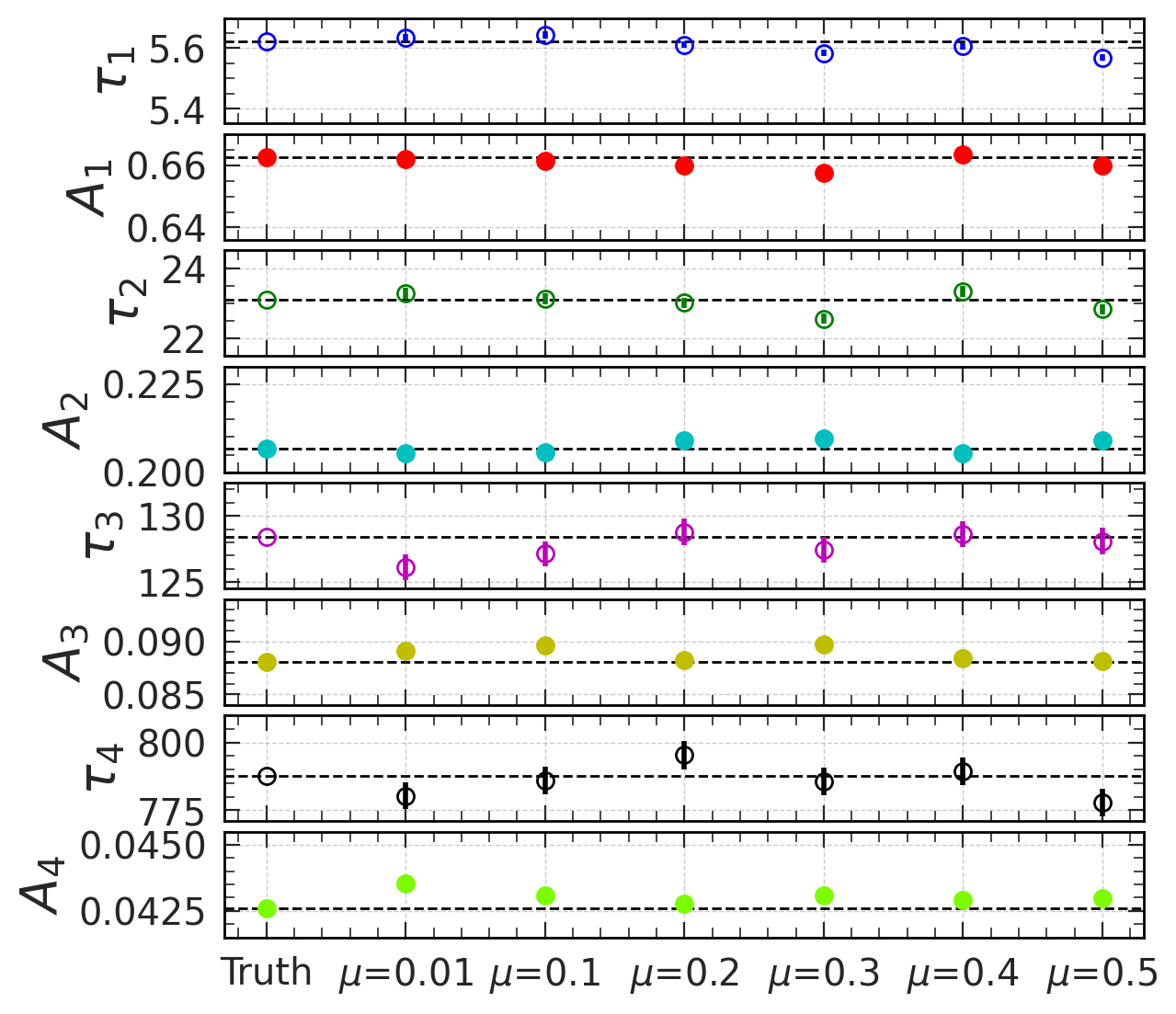}
\qquad
\includegraphics[width=.45\textwidth]{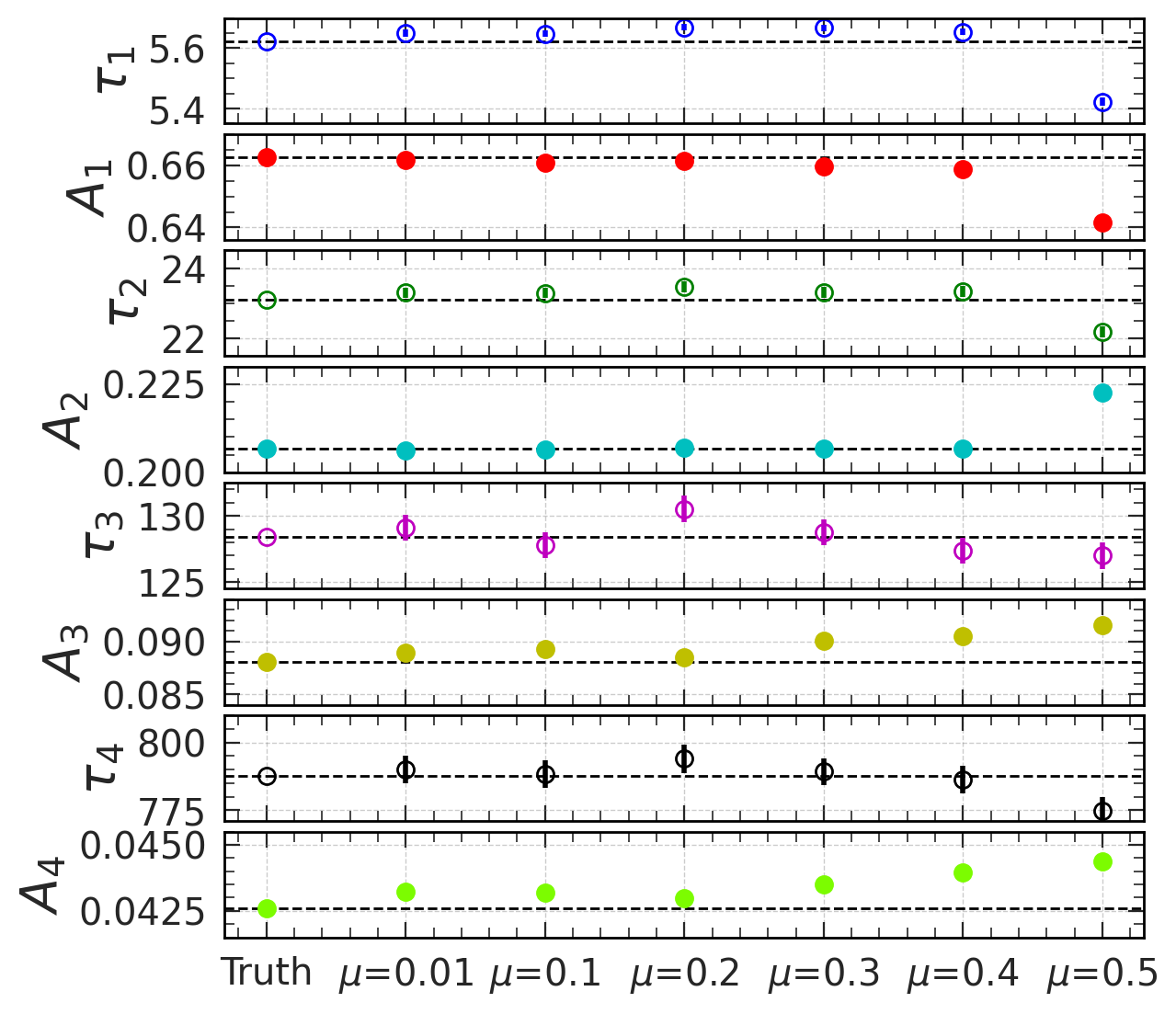}
\caption{Best fit scintillation time parameters as a function of average detected photon number. The left panel displays results excluding the cross-talk effect, whereas the right panel includes this effect. 
\label{fig:best_fit_syst_with_occ}}
\end{figure}

The measured time distribution is used to sample the number of hits within a time window. For each hit, the SPE amplitude is sampled from the measured SPE charge response. A waveform simulation takes into account the SPE pulse shape. A reconstruction algorithm, identical to the one applied to real data, is used to identify the hits. The same cut as outlined in Sec.~\ref{sec:hit_sel} is applied. The results are shown in Figure~\ref{fig:occ_dependence} and \ref{fig:best_fit_syst_with_occ}. Different scenarios with various assumed occupancies are investigated. The results indicate  <2\% shape distortion at <5 ns at an average $\mu=0.5$. No obvious bias on the best fit scintillation time constants are observed.

The impact from cross talk as discussed in Sec.~\ref{sec:hit_sel} is also evaluated, by modeling the cross talk in the toy data. The cross talk waveform template is taken from data. The results are shown in the right panel in Figure~\ref{fig:occ_dependence} and \ref{fig:best_fit_syst_with_occ}. The distortion is slightly larger compared to without adding the cross talk effect, up to 5\%(10\%) at an average $\mu=0.2 (0.5)$. A comprehensive fit to the spectrum reveals minimal deviation when $\mu\leq0.2$ while up to 5\% bias in scintillation time constants with increased $\mu$ to 0.5.

\subsection{Discussion}
\label{sec:discussion}
With the measured time spectrum and extracted time constants, we found ions with Z=1 and Z=2, having an atomic energy of approximately 200 MeV/u,  exhibit a time spectrum similar to that of electrons in the MeV energy range. This finding is different from the typical measurements with MeV-scale neutron sources, where the time response is more similar to alpha particles. This is the first time, with experimental data, we show the  direct time spectrum's dependence  on $dE/dX$ with drastically different particles.

\begin{figure}
\centering
\includegraphics[width=0.6\linewidth]{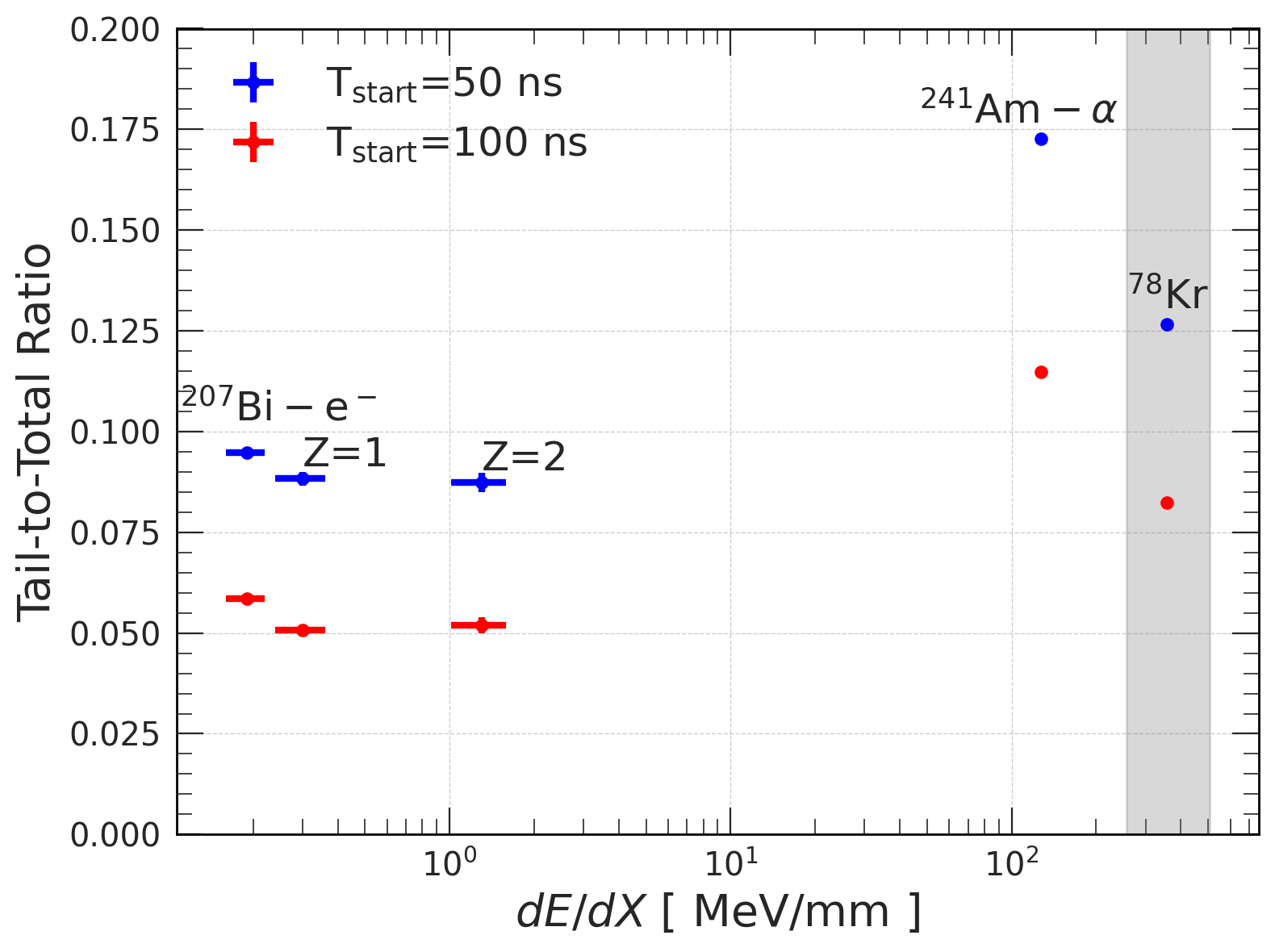}
\caption{\label{fig:ttr_dedx} Scintillation time tail-to-total ratio against $dE/dX$. $dE/dX$ is based on simulation. The value for $^{78}$Kr accounts for the energy deposition from the main track.}
\end{figure}

To represent the shape difference in the context of pulse shape discrimination, we showed the simplest tail-to-total ratio (TTR) for these particles, plotted as a function of $dE/dX$ in Figure~\ref{fig:ttr_dedx}. We need to point out that the exact particle discrimination power also varies with the overall light yield of the experimental setup. The TTR here is calculated using the best fit truth scintillation spectrum. At $dE/dX$ regions below 10 MeV/mm, the TTR from electron, proton, and alpha isotopes are similar. The same type of alpha particle, but with lower energy, shows a TTR value 0.07 higher. This indicates underlying energy deposition density is a main factor for scintillation timing. Further data are needed to fill the gap within the two orders of magnitudes of $dE/dX$.

For the $^{78}$Kr particles, the same GEANT4 simulation described in Sec.~\ref{subsec:beam_data} was employed, incorporating \texttt{G4EmLivermorePhysics} and \texttt{G4EmExtraPhysics} to handle secondary electron production, along with \texttt{G4IonPhysicsPHP} for ion processes. The simulations reveal that the $dE/dX$ for the main tracks of $^{78}$Kr are significantly larger than that of the 5.5 MeV alpha particles from $^{241}$Am. However, approximately 20\% of the Kr energy converts into knock-out electrons. Therefore, at least 20\% of the optical photons are generated by electrons with much lower $dE/dX$. Consequently, the overall time spectrum exhibits a smaller tail compared to that of $\alpha$ particles. A more in-depth study is beyond the scope of this work, and we plan to address it in future research. This kind of behavior has been seen for the energy response, e.g. in \cite{Ahlen1977}, where the quenching of energy is unsaturated for high energy relativistic ions. 

%=================%
\section{Conclusion}
\label{sec:conclusion}

In this study, we investigated the $dE/dX$ dependence of the scintillation time response of LAB-based liquid scintillator using both MeV-scale radioactivity sources and high-energy ions. A key finding from our study is the similarity in time responses between different type of particles with similar $dE/dX$. This result contrasts with typical observations involving MeV-scale neutron sources where proton responses typically close to those of alpha particles. This work shows for the first time with experimental data, a direct dependence of the scintillation time spectrum on $dE/dX$ across a diverse range of particles. This finding underscores the importance of considering particle energy loss characteristics when analyzing scintillation time distributions.

The study also highlighted the complex interactions of high-energy $^{78}$Kr ions, where the $dE/dX$ was significantly larger than that of typical alpha particles. However, the observed time constants revealed fewer long-time components, attributed to secondary particle generation. More than 20\% of the optical photons were generated by secondary electrons with lower $dE/dX$, influencing the overall time spectrum. While our current work provides foundational insights, a comprehensive analysis of these secondary interactions will be critical in the future.

In summary, our findings contribute to a deeper understanding of scintillation time responses, bridging gaps in knowledge regarding particle interactions at different energy scales. Future work will aim to measure the time response for the currently uncovered $dE/dX$ range with proper particles, towards a precision of time response models in scintillation detectors.

%=============================%
\section*{Acknowledgments}
This work was partially supported by the National Natural Science Foundation of China (Grants No. 12125506), by CAS Project for Young Scientists in Basic Research (Grant No. YSBR-099). The authors would like to thank Xiaodong Tang for suggesting using the ion beam and for the support about the beam usage. Also, we thank the HIRFL-CSR accelerator team for their efforts to provide a stable beam condition for the experiment and the onsite help from the crews of External Target Facility. Thanks to Haoqi Lu, Jilei Xu, Zhiming Wang, Guofu Cao, Yong Liu, Jiaxuan Ye, Ziyan Deng and Junyu Shao for the hardware support and suggestions during our detector construction. Thanks to Xilei Sun for the support of the calibration sources.  Thanks to Mingxia Sun and Jack Rolph for optimizing the fitter algorithm.

\clearpage

\bibliographystyle{JHEP}
\bibliography{main}

\end{document}